\documentclass[11pt]{article}

\usepackage[]{graphics,graphicx}

\usepackage{pstricks}
\usepackage{amsthm}
\usepackage{blindtext}

\newcommand{\lb}{\left(}
\newcommand{\rb}{\right)}

\usepackage{setspace} 

\RequirePackage{lineno}

\begin{document} 

 \modulolinenumbers[1]


\begin{center}{\LARGE\bf Dispersive landslide}
\\[7mm]
{{\it Shiva P. Pudasaini}
\\[3mm]
{Technical University of Munich, Chair of Landslide Research}\\
{Arcisstrasse 21, 80333, Munich, Germany}
\\[1mm]
{E-mail: shiva.pudasaini@tum.de}\\[7mm]
}
\end{center}
\noindent
{\bf Abstract:} 
Considering the non-hydrostatic mass flow model (Pudasaini, 2022), here, we derive a novel dispersive wave equation for landslide. The new dispersive wave for landslide recovers the classical dispersive water waves as a special case. We show that the frequency dispersion relation for landslide is inherently different than the classical frequency dispersion for water waves. The wave frequency with dispersion increases non-linearly as a function of the wave number. For dispersive landslide, the wave frequency without dispersion appears to heavily overestimate the dispersive wave frequency for higher wave number. Due to the dispersion term emerging from the non-hydrostatic contribution for landslide, the phase velocity becomes a function of the wave number. This gives rise to the group velocity that is significantly different from the phase velocity, characterizing the dispersive mass flow. The dispersive phase velocity and group velocity decrease non-linearly with the wave number. Yet, the group velocity is substantially lower than the phase velocity. We analytically derive a dispersion number as the ratio between the phase velocity and the group velocity, which measures the deviation of the group velocity from the phase velocity, provides a dynamic scaling between them and summarizes the overall effect of dispersion in the mass flow. The dispersion number for landslide increases rapidly with the wave number, which is in contrast to the dispersion in water waves. With the definition of the effective dispersive lateral stress, we prove the existence of an anti-restoring force in landslide. We reveal the fact that due to the anti-restoring force, landslides are more dispersive than the piano strings. So, the wave dispersion in landslide is fundamentally different than the wave dispersion in the piano string. Our model constitutes a foundation for the wave phenomenon in dispersive mass flows.

\section{Introduction}

Landslides and debris avalanches consist of a mixture of granular materials and the fluid. There have been rapid advancements in modeling such mass movements as shallow flows (Savage and Hutter, 1989; Iverson and Denlinger, 2001; Pitman and Le, 2005; Kuo et al., 2011; Pudasaini, 2012; Pudasaini and Mergili, 2019). 
Classically, modeling geophysical flows is based on the hydrostatic, depth-averaged mass and momentum balance  equations (Pudasaini and Hutter, 2007). 
However, in rapid mass flows down inclined slopes the gravity and the vertical acceleration can have the same order of magnitude effects demanding for the non-hydrostatic model formulation (Denlinger and Iverson, 2004; Castro-Orgaz et al., 2015). 
\\[3mm]
The Boussinesq-type water wave theory is widely used in hydraulics and water wave simulations. Following the work of Boussinesq (1872, 1877), the free surface water flow simulations are generally based on non-hydrostatic depth-averaged models. Fundamental further contributions in including Boussinesq-type non-hydrostatic and dispersive effects in water waves are also due to Serre (1953), Peregrine (1967), Green et al. (1976), and Nwogu (1993). 
However, for shallow granular flows, Denlinger and Iverson (2004) included the effect of nonzero vertical acceleration on depth-averaged  momentum fluxes and stress states  while modeling granular flows across irregular terrains. This was later extended by Castro-Orgaz et al. (2015) resulting in the novel Boussinesq-type theory for granular flows. Yuan et al. (2018) advanced further by presenting a refined and more complete non-hydrostatic shallow granular flow model. 
\\[3mm]
Pudasaini (2022) extended and utilized the above mentioned ideas to the multi-phase mass flow model (Pudasaini and Mergili, 2019) to generate a non-hydrostatic Boussinesq-type gravity wave model for multi-phase mass flows. 
The new non-hydrostatic multi-phase mass flow model includes enhanced gravity and dispersive effects as in the single-phase models by Denlinger and Iverson (2004), Castro-Orgaz et al. (2015) and Yuan et al. (2018). However, the Pudasaini (2022) model further includes interfacial momentum transfers in the non-hydrostatic Boussinesq-type model formulation representing the complex multi-phase nature of mass flow. 
\\[3mm]
Here, we consider the non-hydrostatic multi-phase mass flow model (Pudasaini, 2022) and reduce its complexity to a geometrically two-dimensional landslide motion as a mixture of solid particles and fluid down a slope. Then, we derive a novel dispersive wave equation for the landslide motion. We show that the frequency dispersion relation for landslide is essentially different than the classical frequency dispersion for water waves. As the dispersion originates from the non-hydrostatic contribution for landslide, the phase velocity becomes a function of the wave number, resulting in the significantly different group velocity than the phase velocity. The dispersive group velocity is substantially lower than the phase velocity as both decrease non-linearly with the wave number. Analytically derived dispersion number measures the departure of the group velocity from the phase velocity, and encapsulates the overall effect of dispersion in the landslide wave dynamics. The dispersion number for landslide increases rapidly with the wave number.  Existence of an anti-restoring force in landslide proves that landslides are more dispersive than the piano strings. These are new understanding for the dispersive landslide motions. 

\section{A dispersive wave equation for mass flow}

\subsection{Balance equations for mass flow}

A geometrically two-dimensional motion down a slope is considered. Let $t$ be time, $(x,
z)$ be the coordinates and $\left ( g^x, g^z\right )$ the gravity accelerations along and perpendicular to the slope, respectively. Let, $h$ and $u$ be the
flow depth and the mean flow velocity of the landslide along the slope. Similarly,
$\gamma, \alpha_s, \mu$ be the density ratio between the fluid and the
particles $\left ( \gamma = \rho_f/\rho_s\right )$, volume fraction of the solid
particles (coarse and fine solid particles), and the basal friction coefficient $\left ( \mu =
\tan\delta\right )$, where $\delta$ is the basal friction angle of the solid particles, in
the mixture material. Furthermore, $K$ is the earth pressure coefficient, and $C_{DV}$ is the viscous drag coefficient.
 \\[3mm]
 We start with the non-hydrostatic multi-phase mass flow model (Pudasaini, 2022).
 Reducing the sophistication, we consider a landslide motion as an effectively single-phase mixture of solid particles and fluid down a slope. This leads to
a single mass and
momentum balance equation describing the motion of a landslide (or a mass flow) with the non-hydrostatic contributions as:
\begin{equation}
\frac{\partial h}{\partial t} +  \frac{\partial }{\partial x}\left ( hu\right ) = 0,
\label{Eqn_1}
\end{equation}
  {\small
 \begin{eqnarray}
&&\frac{\partial u}{\partial t} +  u\frac{\partial u}{\partial x}  
+ \left [ \left \{\left( \left( 1-\gamma\right)K + \gamma\right)\alpha_s+\left ( 1-\alpha_s \right )\right\}g^z
+\alpha_s \left\{ \lb \frac{\partial}{\partial t}+u \frac{\partial}{\partial x}\rb {w} + C_{_{DV}} w u \right\}
\right ] \frac{\partial h}{\partial x}
\nonumber\\
&&+ \frac{1}{h}\frac{\partial }{\partial x}\left[\left \{ \alpha_s\left (K -1 \right)+1\right\}\left [\frac{h^3}{12} \left\{ \left ( \frac{\partial u}{\partial x}\right )^2 -\frac{\partial}{\partial x}\frac{\partial u}{\partial t} - u \frac{\partial^2u}{\partial x^2}
-2C_{_{DV}} u \frac{\partial u}{\partial x}
\right\}
+\frac{h^2}{2} \left\{ \lb \frac{\partial}{\partial t}+u \frac{\partial}{\partial x}\rb {w} + C_{_{DV}} w u \right\}   \right]\right]\nonumber\\ 
&&= g^x
  -\mu\alpha_s\left[ \lb 1- \gamma\rb g^z + \left\{ \lb \frac{\partial}{\partial t}+u \frac{\partial}{\partial x}\rb {w} + C_{_{DV}} w u \right\} \right]
- C_{_{DV}} u^2.
\label{Eqn_3}
\end{eqnarray}
}
\hspace{-3mm}
The second term on the left hand side of (\ref{Eqn_3}) describes the advection, while the third term (in the first square bracket) describes the extent of the local deformation that stems from the hydraulic pressure gradient of the free-surface of the landslide in which, $\left ( 1-\alpha_s\right)g^z\partial h/\partial x$ emerges from the hydraulic pressure gradient associated with possible interstitial fluids in the landslide, and the terms associated with ${w}$ are from the enhanced gravity (Pudasaini, 2022). The fourth term on the left hand side (in the second square brackets) are extra contributions in the flux due to the non-hydrostatic contributions. Moreover, the third and fourth terms on the left hand side, and the other terms on the right hand side of (\ref{Eqn_3}) represent all the involved forces. 
The first and second
terms on the right hand side of (\ref{Eqn_3}) are
 the gravity
acceleration, effective Coulomb friction that includes
lubrication $\left ( 1- \gamma\right )$, liquefaction $\left (
\alpha_s\right )$ (because, if there is no or substantially low amount of solid, the mass is fully
liquefied, e.g., lahar flows), the third and fourth terms with $w$ emerge from enhanced gravity, and the fifth term is the viscous drag,
respectively. The term with $1-\gamma$ or $\gamma$ originates from the buoyancy effect. By setting $\gamma = 0$ and $\alpha_s = 1$, we obtain a dry landslide, grain flow, or an avalanche motion. However, we keep $\gamma$ and $\alpha_s$ also to include possible fluid effects in the landslide (mixture). 
Note that for $K = 1$ (which may prevail for extensional flows, Pudasaini and Hutter, 2007), the third term on the left hand side associated with $\partial h/\partial x$ simplifies drastically, because $\left \{ \left( \left( 1-\gamma\right)K + \gamma\right)\alpha_s+\left ( 1-\alpha_s \right )\right \}$ becomes unity.
 So, the isotropic assumption (i.e., $K = 1$) loses some important information about the solid content and the buoyancy effect in the mixture.
 Furthermore, $\displaystyle{w = - \frac{h}{2}\frac{\partial u}{\partial x}}$ is the mean slope normal velocity (Yuan et al., 2018; Pudasaini, 2022).

\subsection{Linearized mass and momentum balance equations}

 We linearize (\ref{Eqn_1}) and (\ref{Eqn_3}) with $h = H + \tilde h$, where $H$ is the background (mean) material depth on which the amplitude $\tilde h$ is defined. For simplicity, the tildes are discarded from the resulting equations. 
 Then, we obtain the linearized mass and momentum equations as:
 \begin{equation}
\frac{\partial h}{\partial t} +  H\frac{\partial u}{\partial x} = 0,
\label{Eqn_4}
\end{equation}
{\small
\begin{equation}
\frac{\partial u}{\partial t} 
+ \left [ \left( \left( 1-\gamma\right)K + \gamma\right)\alpha_s+\left ( 1-\alpha_s \right )\right ]g^z
\frac{\partial h}{\partial x}
-\frac{H^2}{3}\left [ \alpha_s\left (K -1 \right)+1\right ]\frac{\partial ^2}{\partial x^2}\frac{\partial u}{\partial t} = g^x-(1-\gamma)\alpha_s\mu g^z
+\displaystyle{\frac{1}{2}\mu\alpha_s H\frac{\partial}{\partial t}\frac{\partial u}{\partial x}},
\label{Eqn_5}
\end{equation}
}
\hspace{-3mm}
where, out of 
$\displaystyle{\frac{H^2}{3}\left [ \alpha_s\left (K -1 \right)+1\right ]\frac{\partial ^2}{\partial x^2}\frac{\partial u}{\partial t}}$ the factors $\displaystyle{\frac{1}{4}}$ and $\displaystyle{\frac{1}{12}}$ stem from the enhanced gravity (or hydraulic pressure gradients) and dispersions, respectively.
We note that, (\ref{Eqn_5}) extends the Peregrine (1967) dispersive system (Khakimzyanov et al., 2020a,b) from water waves to mixture debris waves. 

\subsection{A novel dispersive wave equation for landslide}

Now, utilizing (\ref{Eqn_4}), the third terms on both sides of (\ref{Eqn_5}) can be written in terms of $h$, and the resulting momentum equation yields:
\begin{equation}
\frac{\partial u}{\partial t}   
+\left [ \left( \left( 1-\gamma\right)K + \gamma\right)\alpha_s+\left ( 1-\alpha_s \right )\right ]g^z
\frac{\partial h}{\partial x}
+\frac{H}{3}\left [ \alpha_s\left (K -1 \right)+1\right ]\frac{\partial}{\partial x}\frac{\partial^2 h}{\partial t^2} = g^x-(1-\gamma)\alpha_s\mu g^z
-\displaystyle{\frac{1}{2}\mu\alpha_s\frac{\partial^2 h}{\partial t^2}}.
\label{Eqn_6}
\end{equation}
With the help of (\ref{Eqn_4}), $u$ can be removed from (\ref{Eqn_6}). For this, differentiate (\ref{Eqn_4}) with respect to $t$, and (\ref{Eqn_6}) with respect to $x$. Then, eliminating $u$ from (\ref{Eqn_6}), we ordain a novel dispersive wave equation for landslide:
\begin{equation}
\frac{\partial^2 h}{\partial t^2} = \mathcal T \frac{\partial^2 h}{\partial x^2} + \mathcal D \frac{\partial^2}{\partial x^2}\frac{\partial^2 h}{\partial t^2} 
+\mathcal I \frac{\partial^2 }{\partial t^2}\frac{\partial h}{\partial x},
\label{Eqn_7}
\end{equation}
where $\mathcal T, \mathcal D$, and $\mathcal I$ are the involved physical parameters given by 
$\displaystyle{\mathcal T = 
\left [ \left( \left( 1-\gamma\right)K + \gamma\right)\alpha_s+\left ( 1-\alpha_s \right )\right ]
g^z H,}$ \\
        $\displaystyle{\mathcal D = \left [ \alpha_s\left (K -1 \right ) + 1\right ]  \frac{H^2}{3},}$
        and
        $\displaystyle{\mathcal I =\displaystyle{\mu\alpha_s \frac{H}{2}}}$, respectively.
     In  (\ref{Eqn_7}),  $\mathcal T$ is the effective lateral stress (per unit density). For the reasons explained below, we call $\mathcal D$ the dispersion parameter.
        Here, $\mathcal D$ characterizes the non-hydrostatic contribution, and the term associated with $\mathcal I$ emerged due to the effect of enhanced gravity in the source. 
        Note that, for a variable slope, different additional forcing terms would appear in (\ref{Eqn_7}), which have been neglected for now for simplicity. 
        Equation (\ref{Eqn_7}) is a complex dispersive partial differential equation for landslide.
  For a relatively less dense flow (low particle concentration) with lower friction, and/or a relatively slowly varying flow surface, the term with $\mathcal I$ may be ignored. If not, this can be revived.  
  Equation (\ref{Eqn_7}) takes the simple classical wave equation when the terms with $\mathcal D$ and  $\mathcal I$ are ignored, for which $\sqrt{\mathcal T}$ is the wave speed for the debis motion. To explore the first order effects of the dispersive phenomena in the mixture mass flow, in what follows, for simplicity, we disregard the influence of the term associated with $\mathcal I$.

  \section{Dispersion in non-hydrostatic mass flow}
  
\subsection{The dispersion relation}

Assume a plain wave of the form:
\begin{equation}
h = h_0\exp\left[i(kx - \omega t)\right],
\label{Eqn_8}
\end{equation}
where $h_0 = h(0,0)$, and $\omega$ and $k$ are the wave frequency and the wave number ($\sim$ reciprocal of the wave length).
Applying  (\ref{Eqn_8}) in to  (\ref{Eqn_7}) results in:
\begin{equation}
\omega^2  =  \mathcal T k^2 - \mathcal D \omega^2 k^2, 
\label{Eqn_9}
\end{equation}
 We write (\ref{Eqn_9}) in the form
\begin{equation}
\omega^2  =  \frac{\mathcal T k^2} {1 + \mathcal D k^2}. 
\label{Eqn_10}
\end{equation}
Equation (\ref{Eqn_10}) is the frequency dispersion relation (linking frequency and wave number) to our model for mass flow, which is different than the classical linear frequency dispersion for Boussinesq water wave equations (Dingemans, 1997). Note that, in (\ref{Eqn_10}), $\mathcal D$ is proportional to $H^2$. So, $\mathcal D k^2 \sim H^2k^2$, which contains the relative wave number $Hk$.

\subsection{The phase velocity}

The landslide phase velocity (speed) $C_p$ is defined as $C_p = \omega/k$, which, from (\ref{Eqn_10}), takes the form

\begin{equation}
C_p  =  \frac{\omega}{k}= \pm \sqrt{\frac{\mathcal T } {1 + \mathcal D k^2}}.
\label{Eqn_11}
\end{equation}
So, for the non-hydrostatic mass flow, the phase velocity is not a constant but is a function of the wave number. Due to the non-zero positive dispersion parameter $\mathcal D$, (\ref{Eqn_11}) gives rise to the group velocity that is different from the phase velocity. Also note that, $\mathcal T$ has the dimension of m$^2$s$^{-2}$ and $\mathcal D k^2$ is dimensionless. Thus, $C_p$ has the dimension of ms$^{-1}$, the velocity.
 
 \subsection{The group velocity}
 
 The landslide group velocity is denoted by $C_g$ and is defined as
 \begin{equation}
C_g  = \frac{\partial \omega}{\partial k},
\label{Eqn_14}
\end{equation}
which is the measure of the rate of change of the wave frequency as a function of the wave number.
From the phase velocity (\ref{Eqn_11}), we obtain the group velocity:
\begin{equation}
C_g  = \frac{\partial \omega}{\partial k} = \frac{\partial}{\partial k}\left ( k C_p\right) 
=\frac{1}{\mathcal T}\left ( \frac{\mathcal T}{1+\mathcal D k^2}\right )^{3/2} = \left ( \frac{1}{1+\mathcal D k^2}\right) C_p.
\label{Eqn_15}
\end{equation}

\subsection{The dispersion number}

Equation (\ref{Eqn_15}) reveals a strikingly impressive relation between the phase velocity and the group velocity. There exists a function $\mathcal D_n$ such that it defines a mapping between the phase velocity and the group velocity given by the relation:
 \begin{equation}
C_p  = C_g \mathcal D_n, \,\,\,\,\, \mathcal D_n(k) = {1+\mathcal D k^2}.
\label{Eqn_15_Map}
\end{equation}
This function can be written as 
\begin{equation}
 \mathcal D_n = \frac{C_p}{C_g}.
\label{Eqn_15_Dispersion_Nr}
\end{equation}
As for $C_p$, $C_g$ has the dimension of velocity. So, $\mathcal D_n$, as the ratio between the phase velocity and the group velocity, is a dimensionless number. 
 We call $\mathcal D_n$ the dispersion number that measures the deviation of the group velocity from the phase velocity. 
 As indicated by (\ref{Eqn_15_Map}), for the problem under consideration,  $\mathcal D_n$ is a stretching function of the wave number and is bounded from below by unity, i.e., $\mathcal D_n \ge 1$. This means, $\mathcal D_n$ provides a dynamic scaling between $C_g$ and $C_p$. It shows that $C_g \le C_p$, which is opposite to the dispersive wave in a piano string (Podlesak and Lee, 1988; Gracia and Sanz-Perela, 2017). Furthermore, for our problem, the phase and the group velocity have the same direction, but different speeds.
 Moreover, the wave becomes non-dispersive if the term associated with $\mathcal D$ can be ignored for which $C_g \to C_p$, consequently $\mathcal D_n = 1$, as for the ideal string, or the sound wave in a room, which are non-dispersive. However, for piano string, $\mathcal D_n$ takes the form $\mathcal D_n = \lb 1 + \mathcal D k^2\rb/\lb 1 + 2\mathcal D k^2\rb $, $\mathcal D$ here appropriately corresponds to the physical quantity for piano. It means, for piano string the dispersion number
 decreases as a function of the wave number $k$ from its maximum 1 (as $k\to 0$) to minimum $1/2$ (as $k$ is sufficiently large). So, the wave dispersion for landslide is fundamentally different than the wave dispersion in the piano string. In other words, landslides can be more dispersive than the piano strings.
 \\[3mm]
 In fact, the dispersion number plays a dominant role as all the relevant quantities $\omega, C_p$ and $C_g$ are expressed in terms of $\mathcal D_n$. Importantly, once we know $\mathcal T$ and $\mathcal D_n$, the wave frequency, phase and group velocities are known, because, usually, $\mathcal T$ is a parameter, and $\mathcal D_n$ varies as a function of the wave number.

 \section{Results and analyses of dispersive landslides}
 
 Here, we manifest the contribution of dispersion on the wave motion in mass flow. Unless otherwise stated, following the general values from the literature (Pudasaini and Hutter, 2007; Pudasaini and Mergili, 2019), the material parameters are chosen as follows: the earth pressure coefficient $K = 0.9$ (main downslope extensional motion), the volume fraction of solid in the mixture material $\alpha_s = 0.65$, the buoyancy (lubrication) parameter $\gamma = 1100/2900$ (ratio between the true fluid and the solid densities in the mixture), $g^z = g \cos\zeta$ ($g = 9.81, \zeta = 45^\circ$, the gravitational constant and the slope angle), and the mean material depth $H = 0.5$ m, respectively. Here, $K$ represents the frictional behavior of the material (lower in extension, higher in compression), $\gamma$ the frictional weakening due to the possible presence of the fluid, and $\alpha_s$ characterizes the liquefaction in the mixture material, because as $\alpha_s \to 0$ the mixture is fully liquefied (Pudasaini and Krautblatter, 2021, 2022). So, $\alpha_s, K$ and $\gamma$ together explain the behavior of the granular (debris) material in the mixture. These give the values of $\mathcal T$ and $\mathcal D$ in (\ref{Eqn_11}) and (\ref{Eqn_15}) of about 13.31 and 1.25, respectively. 
 
 \subsection{The wave frequency}
 
\begin{figure}[t!]
\begin{center}
  \includegraphics[width=15cm]{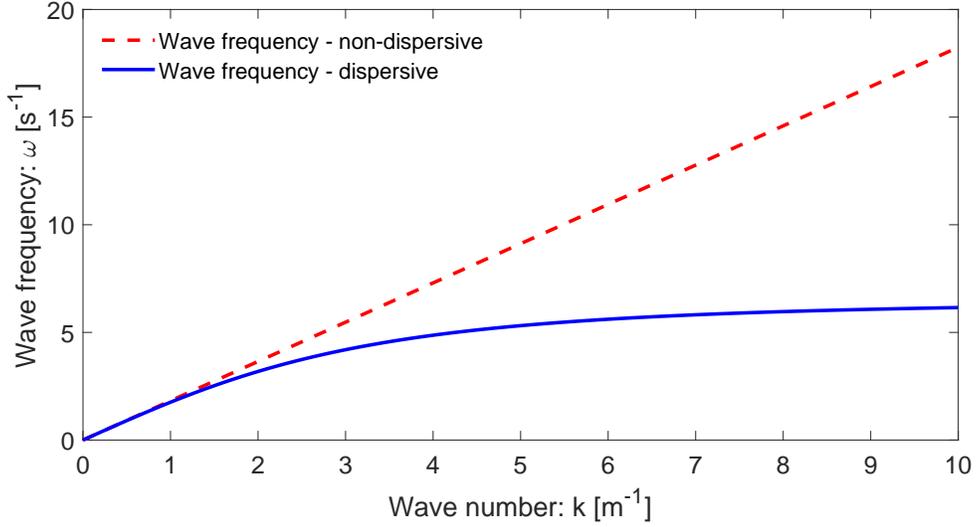}
  \end{center}
  \caption[]{The wave frequency as a function of wave number given by (\ref{Eqn_11}). The wave frequency without and with dispersion are fundamentally different, and differ largely for higher wave number.}
  \label{Fig_1}
\end{figure}

 Figure \ref{Fig_1} displays the wave frequency as a function of the wave number, $\omega = k C_p$, as given by the relation (\ref{Eqn_11}). While the wave frequency increases linearly with the wave number without dispersion (that can be realized by setting $\mathcal D = 0$), the wave frequency with dispersion (including the term associated with $\mathcal D$ that can be realized with $\mathcal D \neq 0$) increases non-linearly as a function of the wave number. For small wave number both wave frequencies are similar, however, for large wave numbers, the difference is large. Furthermore, in general, the wave frequency without dispersion is much higher than the same with dispersion. When in reality the waves are dispersive, the wave frequency without dispersion appears to heavily overestimate the dispersive wave frequency for higher wave number. 

\subsection{The phase velocity and group velocity}

\begin{figure}[t!]
\begin{center}
  \includegraphics[width=15cm]{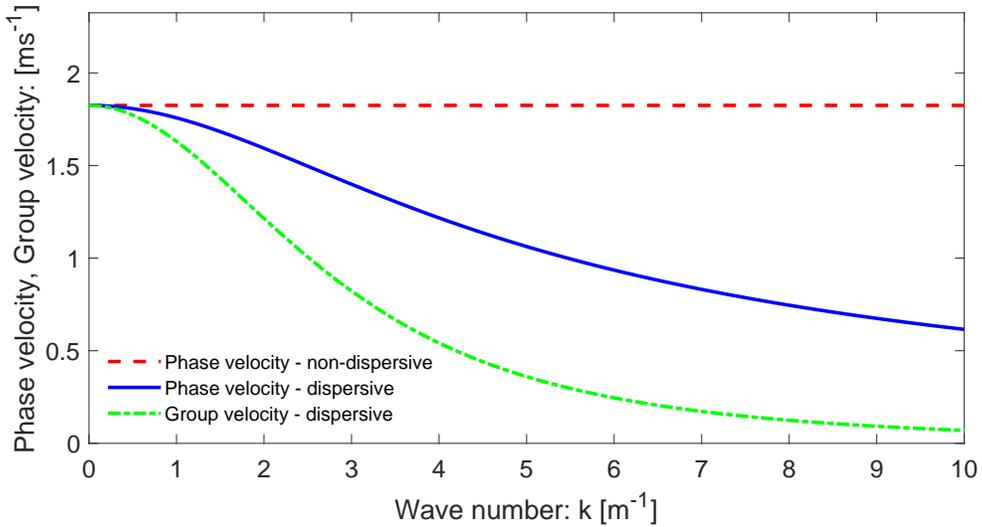}
  \end{center}
  \caption[]{The phase velocity and group velocity as functions of the wave number given by (\ref{Eqn_11}) and (\ref{Eqn_15}). The non-dispersive phase velocity is constant. The dispersive phase velocity and group velocity decrease non-linearly with the wave number. The dispersive group velocity is the lowest among the three.}
  \label{Fig_2}
\end{figure}
The phase velocity and group velocity are technically important quantities as they provide the information of the motion of individual wave crest and energy transport of the modulated wave packet.
The phase velocity without and with dispersion, and the group velocity as given by (\ref{Eqn_11}) and (\ref{Eqn_15}), respectively, are shown in Fig. \ref{Fig_2}. By definition, the non-dispersive phase velocity is a constant. However, the dispersive phase velocity decreases non-linearly as the wave number increases.  Moreover, the group velocity further decreases non-linearly as the wave number increases. Importantly, with dispersion, all three behave fundamentally differently. This, in fact, is the manifestation of dispersion. Without dispersion, all three would be the same with a constant value, the non-dispersive phase velocity. 

\subsection{The dispersion number}

\begin{figure}
\begin{center}
  \includegraphics[width=15cm]{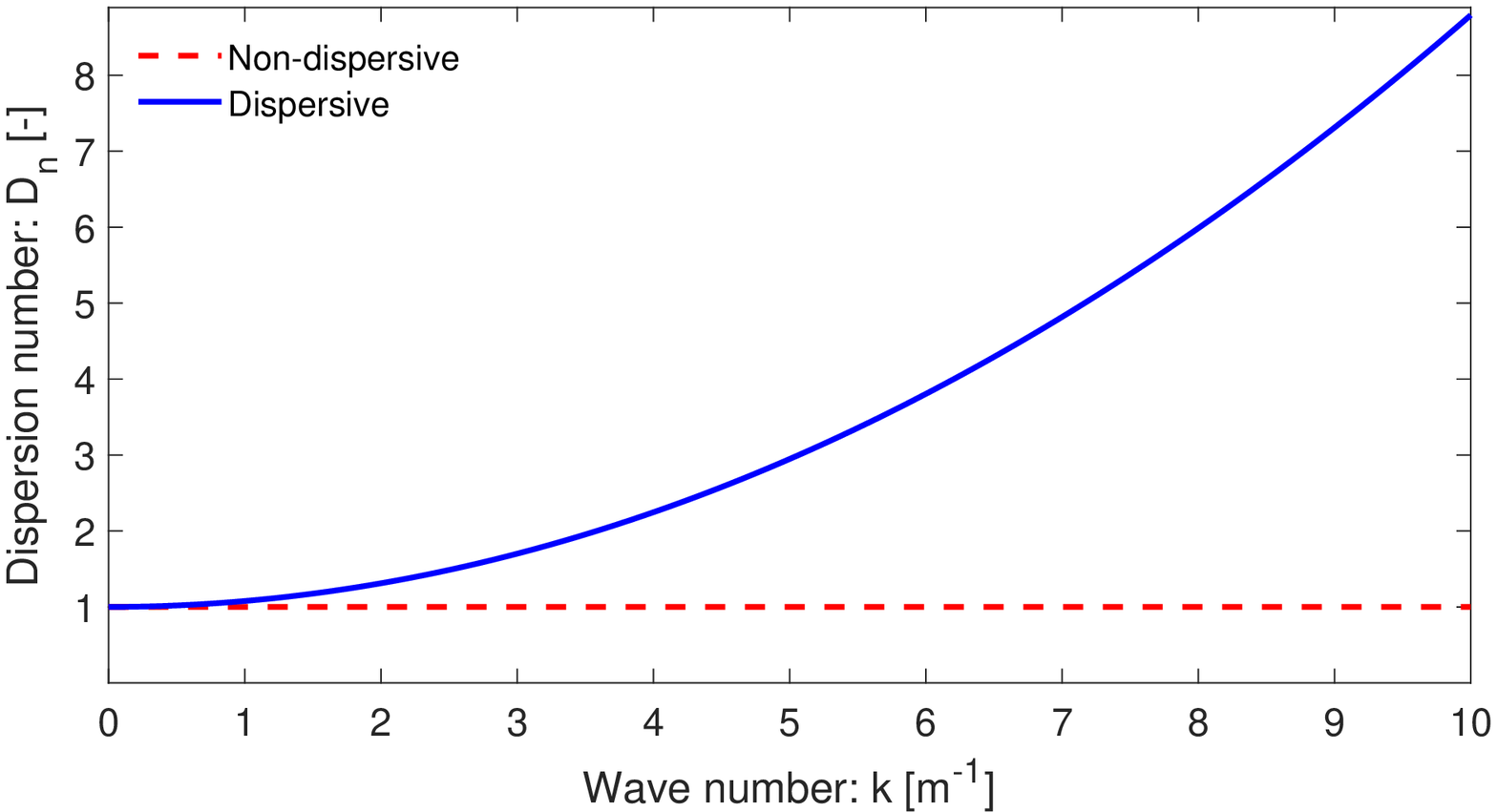}
  \end{center}
  \caption{The dispersion number $\mathcal D_n$ as a function of the wave number $k$ given by (\ref{Eqn_15_Map}). Also shown is the reference when the dispersion is absent.}
  \label{Fig_3}
\end{figure}
The dispersion number $\mathcal D_n$ given in (\ref{Eqn_15_Dispersion_Nr}) is presented in Fig. \ref{Fig_3}. It shows that the dispersion number increases rapidly as the wave number increases. This resulted due to the stretching of $\mathcal D_n$ as given in (\ref{Eqn_15_Map}), and summarizes the overall effect of dispersion in the wave dynamics in mass flow.

\subsection{Influence of parameters}

\begin{figure}[t!]
\begin{center}
  \includegraphics[width=15cm]{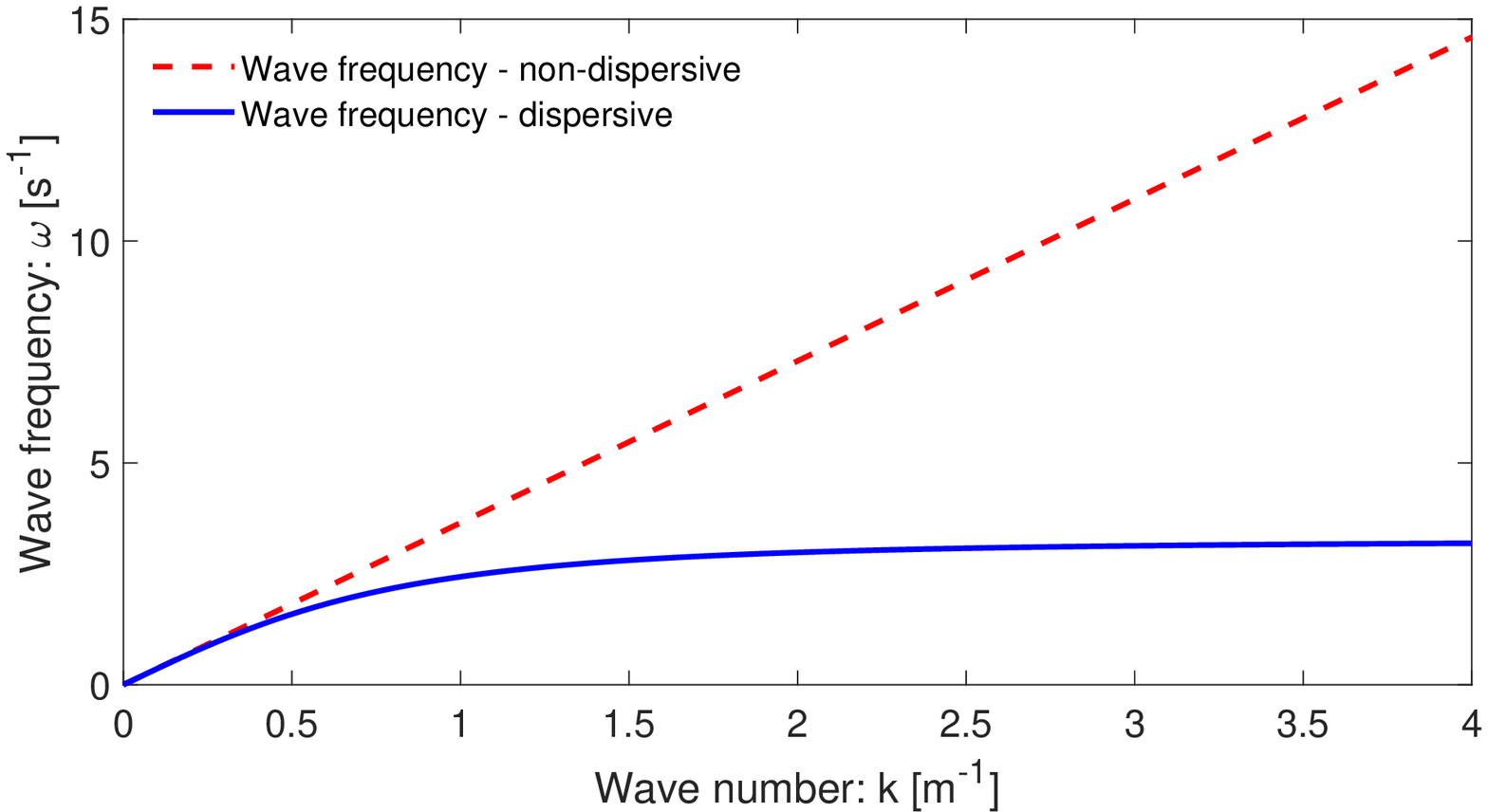}
  \end{center}
  \caption[]{The wave frequency as a function of wave number as in Fig. \ref{Fig_1}, but now with $H = 2$. The wave frequency without and with dispersion are fundamentally different, and differ largely for higher wave number, more than in Fig. \ref{Fig_1}.}
  \label{Fig_1_1}
\end{figure}
\begin{figure}[t!]
\begin{center}
  \includegraphics[width=15cm]{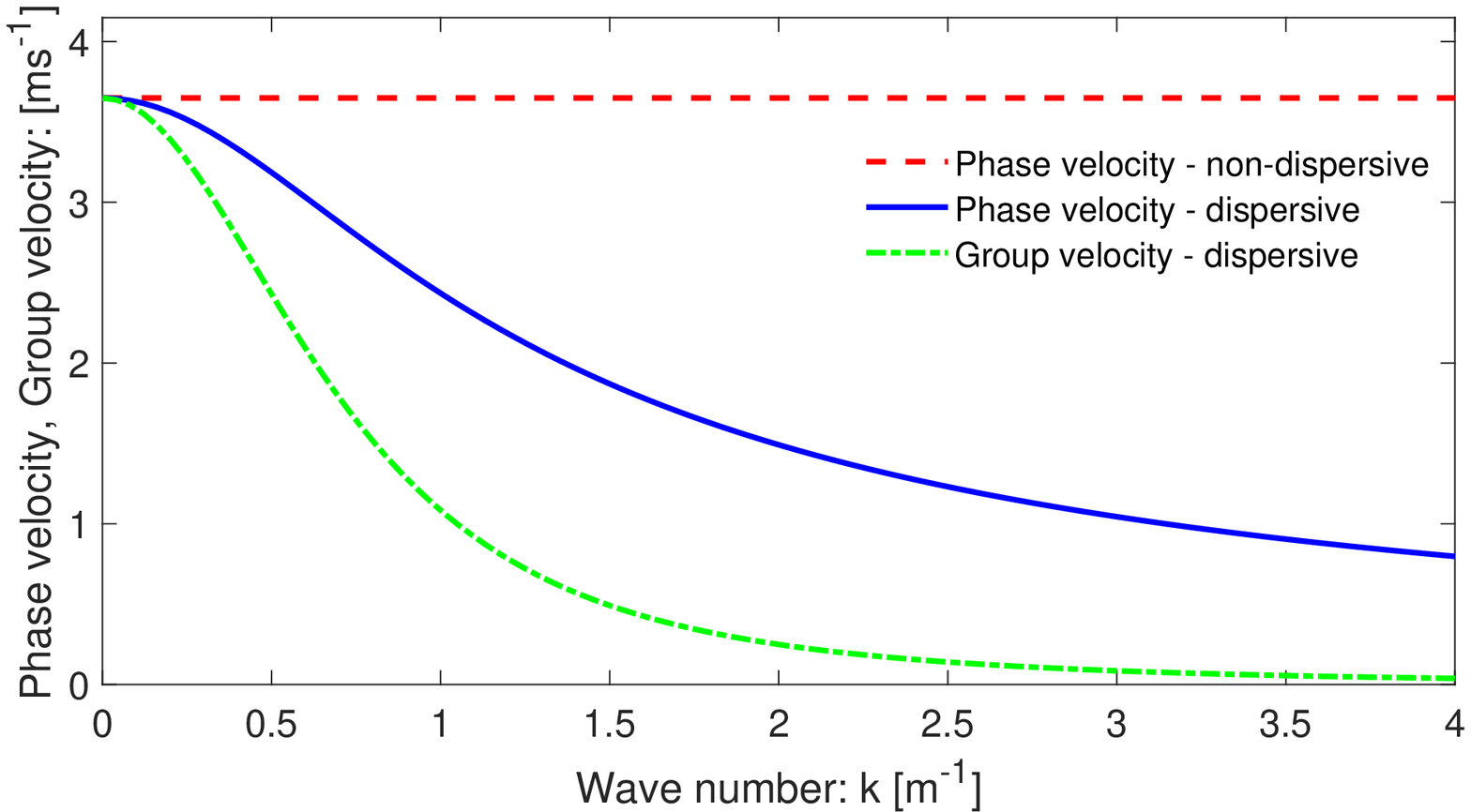}
  \end{center}
  \caption[]{The phase velocity and group velocity as functions of wave number as in Fig. \ref{Fig_2}, but now with $H = 2$. The non-dispersive phase velocity is constant. The dispersive phase velocity and group velocity decrease non-linearly with the wave number, faster than in Fig. \ref{Fig_1}. The dispersive group velocity is the lowest among the three.}
  \label{Fig_2_1}
\end{figure}
\begin{figure}[t!]
\begin{center}
  \includegraphics[width=15cm]{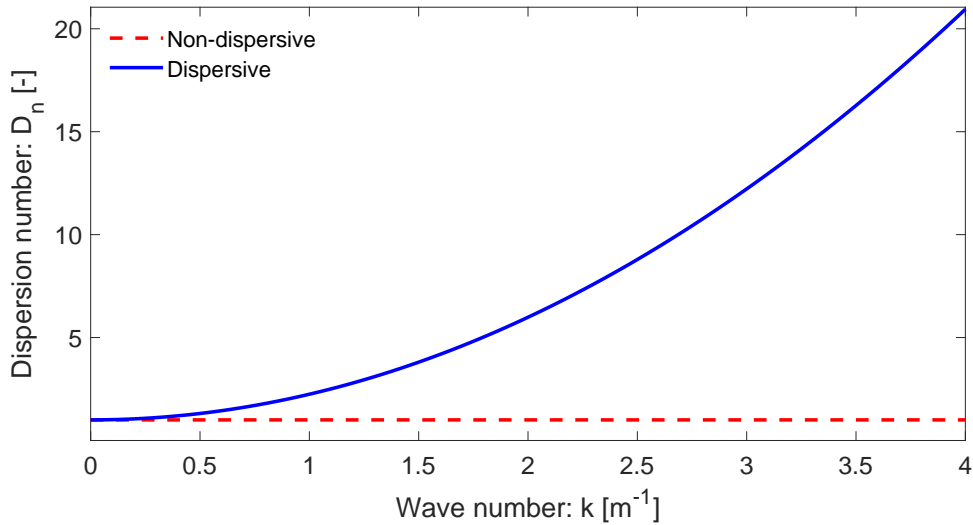}
  \end{center}
  \caption[]{The dispersion number $\mathcal D_n$ as a function of the wave number $k$ as in Fig. \ref{Fig_3}, but now with $H = 2$, which now increases more rapidly than in Fig. \ref{Fig_3}. Also shown is the reference when the dispersion is absent.}
  \label{Fig_3_1}
\end{figure}

The wave frequency, phase and group velocities, and the dispersion number, $\omega$, $C_p$, $C_g$ and $\mathcal D_n$, all depend collectively on the effective lateral stress $\mathcal T$ and the dispersion parameter $\mathcal D$. However, explicitly, they depend either linearly or non linearly on the solid volume fraction $\alpha_s$, the earth pressure coefficient $K$, buoyancy or lubrication effect $\gamma$, the channel slope $\zeta$, and the mean material depth $H$. The detailed analysis can be carried out based on all these parameters. Particularly important are $\alpha_s, K$ and $\gamma$ as they carry crucial physical information of the solid particles and the fluid in the mixture. Nevertheless, as seen from the representations and definitions of $\omega$, $C_p$, $C_g$ and $\mathcal D_n$, $H$ plays a rather key role in determining the wave frequency, phase and group velocities, and the dispersion number, because  $\mathcal D$ varies quadratically with $H$. So, here, we only focus on $H$. We increase its value from 0.5 to 2.  The results are presented in Fig. \ref{Fig_1_1}, Fig. \ref{Fig_2_1} and Fig. \ref{Fig_3_1} for the wave frequency, the phase and group velocities and the dispersion number, respectively. Comparing these figures with their counterparts, Fig. \ref{Fig_1}, Fig. \ref{Fig_2} and Fig. \ref{Fig_3}, it is evident that the wave frequency, phase and group velocities all decrease strongly with the increased $H$ value and saturate much earlier with their lower values within the domain of smaller wave number $k$. However, the dispersion number increases rapidly with the increased $H$ value. This is also what the structures of these variables tell us from their analytical representations, because all of $\omega$, $C_p$, and $C_g$ are somehow inversely related with $\mathcal D$, but, $\mathcal D_n$ is linearly related with $\mathcal D$. However, note that the non-dispersive phase velocity is now significantly higher than in the previous figure, and the rate at which the phase and group velocities decrease is much higher than the same in the previous figure. This also resulted in the rapid increase of dispersion number than in the previous figure for higher mean material depth. 
 
 \subsection{Comparison with the surface water wave}
 
\begin{figure}[t!]
\begin{center}
  \includegraphics[width=15cm]{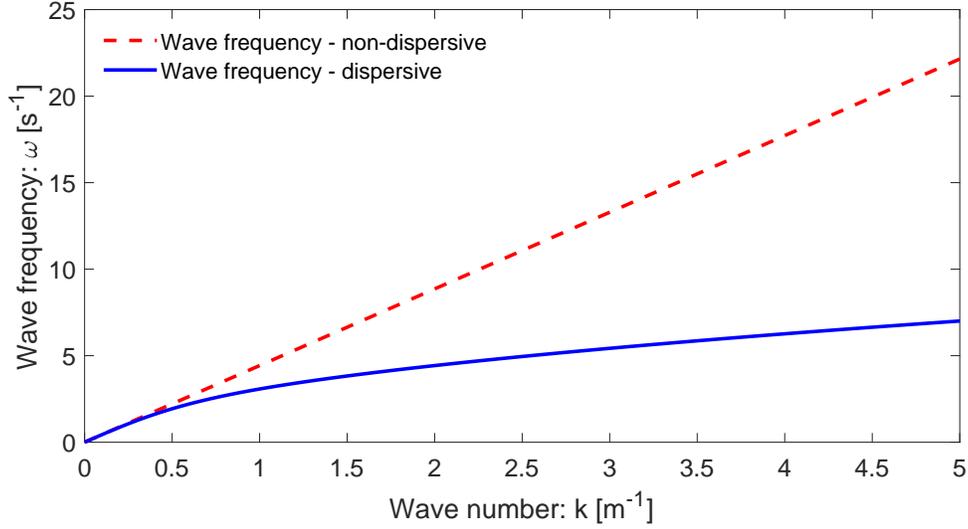}
  \end{center}
  \caption[]{The wave frequency as a function of wave number for surface water waves for intermediate water depth.}
  \label{Fig_1_2}
\end{figure}

\begin{figure}[t!]
\begin{center}
  \includegraphics[width=15cm]{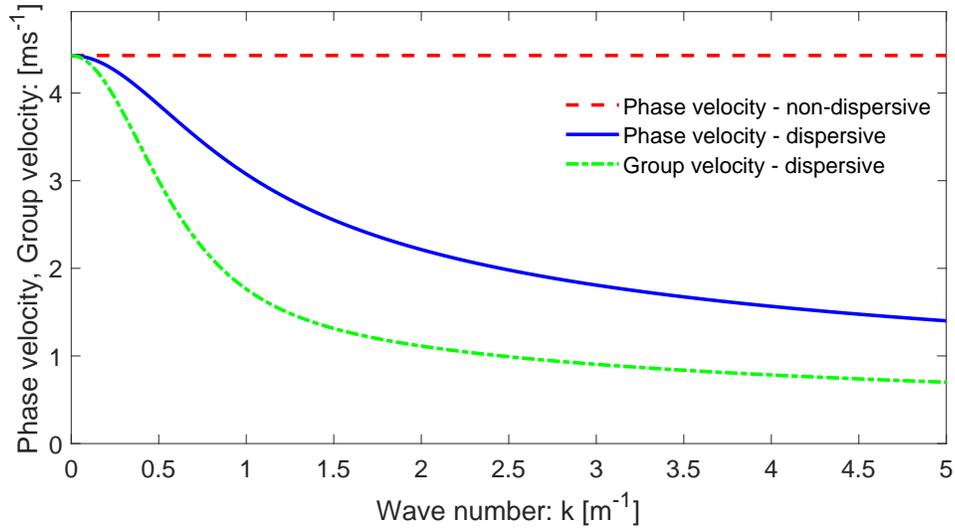}
  \end{center}
  \caption[]{The phase velocity and group velocity as functions of wave number for surface water waves for intermediate water depth.}
  \label{Fig_2_2}
\end{figure}

\begin{figure}[t!]
\begin{center}
  \includegraphics[width=15cm]{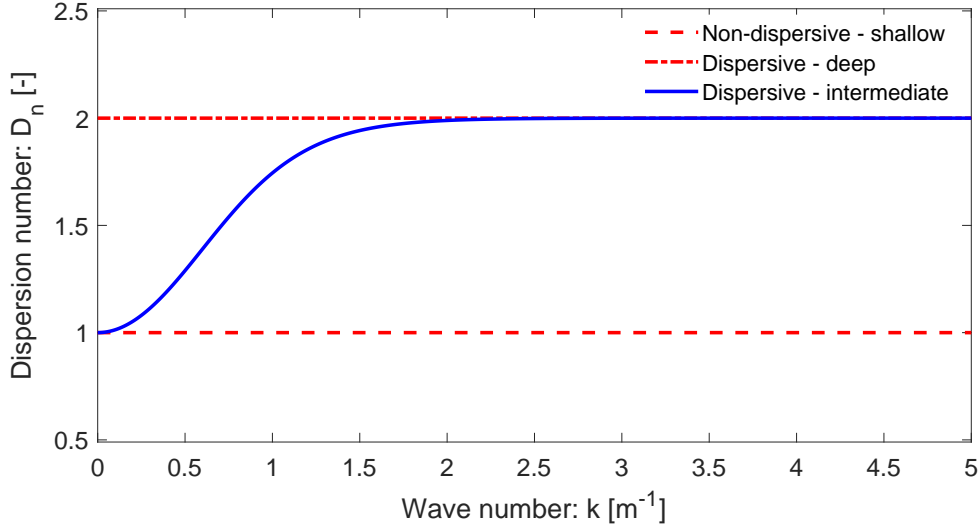}
  \end{center}
  \caption[]{The dispersion number as a function of the wave number for surface water waves for intermediate water depth.}
  \label{Fig_3_2}
\end{figure}

 The classical shallow water surface waves are non-dispersive. For deep water surface waves, the phase velocity is $C_p = \sqrt{g/k}$ and the group velocity is one half of the phase velocity, $C_g = 0.5 C_p$ and are independent of the fluid depth. This is not relevant for us for the present consideration. So, with respect to the dispersion relation, the intermediate fluid depth is relevant here. The phase velocity for the intermediate water depth is $\displaystyle{C_p = \sqrt{\frac{g}{k}\tanh(Hk)}}$, while the group velocity is $\displaystyle{C_g = \frac{1}{2}\left [ 1+ \frac{2Hk}{\sinh(2Hk)}\right ]C_p}$, see, e.g., Dingemans (1997). Dispersion relations for the water waves with the intermediate depth are presented in Fig. \ref{Fig_1_2} for the wave frequency, in Fig. \ref{Fig_2_2} for the phase and group velocity, and in Fig. \ref{Fig_3_2} for the wave number, respectively, with the fluid depth $H = 2$ (chosen this way for the comparison reason).
 Compared with the corresponding figures Fig. \ref{Fig_1_1}, Fig. \ref{Fig_2_1} and Fig. \ref{Fig_3_1}, we observe that the new dispersion relations derived in Section 4 behave fundamentally differently than the dispersion relations for the water waves with depth. Particularly interesting is the dispersion number. While the new dispersion number derived here for mass flow increases continuously as a quadratic function of the wave number with its minimum value of unity, the dispersion number for the water waves also begins at its minimum value of unity, then hyperly rises up, but then asymptotically approaches the dispersion number (two) of the deep water waves already at about wave number $k \ge 2$.  

 \section{Discussion}
 
 \subsection{General aspects of the landslide dispersion relation}

 We observed the following important physical phenomena from the landslide dispersion relation presented in Section 4.
 
\begin{itemize}

\item As $C_p$ depends on the wave number $k$, the wave under consideration is strongly dispersive.  

\item The new dispersion relation includes many different physical parameters and mechanical responses.

\item The usual shallow water wave ($\alpha_s = 0, K = 1$, and the term with $\mathcal D$ can be neglected) is a special case: $C_p^w = \sqrt{g^zH}$.

\item Classical debris-avalanche motion is a special case when the non-hydrostatic contribution is neglected, i.e., $C_p^d = \sqrt{\mathcal T}$, which is the wave speed, and can be written in alternative form as:
{\small
\begin{equation}
C_p^d  =  \sqrt{\mathcal T }= \sqrt{{\left [ \left( \left( 1-\gamma\right)K + \gamma\right)\alpha_s+\left ( 1-\alpha_s \right )\right ]g^z H}}
=\sqrt{{\frac{1}{\rho}\left [ \left( \left( 1-\gamma\right)K + \gamma\right)\alpha_s+\left ( 1-\alpha_s \right )\right ]  \rho g^z H}} = \sqrt{\frac{T}{\rho}},
\label{Eqn_12}
\end{equation}
}
\hspace{-2.5mm}
where, $\rho$ is the mixture bulk density and $T = \left [ \left( \left( 1-\gamma\right)K + \gamma\right)\alpha_s+\left ( 1-\alpha_s \right )\right ] \rho g^z H$ is the lateral stress (compressive for mass flow). So, without the dispersion (non-hydrostatic) contribution, the square of the phase velocity is the ratio between the lateral compressive stress and the material density. Which is similar to the phase velocity for a string in which the ratio is between the tension and the density. 
 Thus, for the mass flow, the compressive stress involves gravity, particle concentration, buoyancy, the earth pressure coefficient and the mean depth of the debris material, which for the classical shallow water is only related to gravity and the mean water depth. 
There are other important aspects: Phase velocity is high for compressional flows (for which $K > 1$), and low for dilational flows ($K < 1$). Similarly, phase velocity is high for pure granular flow ($\alpha_s = 1, \gamma = 0$), and reduces for the particle fluid mixture flows, with its minimum for fully buoyant flows or when the particle concentration vanishes, turning it in to the pure fluid flow. These physical mechanisms are consistent with the strength of material.   

\item Classical shallow water and debris flow models are non-dispersive.

\item {\bf Dispersive lateral stress and dispersion intensity:} Consider $C_p$ in (\ref{Eqn_11}) in its full form involving $\mathcal D$:
\begin{equation}
C_p  =  \sqrt{\frac{\mathcal T } {1 + \mathcal D k^2}}
=\sqrt{\frac{ \left [ \left( \left( 1-\gamma\right)K + \gamma\right)\alpha_s+\left ( 1-\alpha_s \right )\right ] \rho g^z H}{\left({1 + \mathcal D k^2}\right)\rho}} = \sqrt{\frac{T}{\left({1 + \mathcal D k^2}\right)}\frac{1}{\rho}}
= \sqrt{\frac{T_e}{\rho}},
\label{Eqn_13}
\end{equation}
where $T_e = T_e (k) = T/\left({1 + \mathcal D k^2}\right)$. We call $T_e$ the effective dispersive lateral stress. This means that as the wave number increases, the effective dispersive stress decreases, however without changing the material density. This ultimately decreases the phase velocity, as in the reduced phase speed in string, but, with less tension.   Importantly, the dispersion relation emerges due to the dispersion parameter $\mathcal D$ which varies linearly with the solid particle concentration $\alpha_s$ and the lateral pressure coefficient $K$, and quadratically with the depth $H$. So, the dispersion intensity increases linearly with $\alpha_s$ and $K$, and quadratically with $H$ and $k$. Therefore, dispersion is strong for relatively thick flows, and for large wave number. 

\item  {\bf The anti-restoring force in landslide:} $\mathcal D$ in the denominator in $C_p$, i.e., in $\left (1+ \mathcal D k^2 \right)$, generates the dispersive wave. This, in our consideration, originates from the acceleration of the debris material in the slope normal direction (including drags and virtual mass forces in real mixture where the relative acceleration between particle and fluid is not negligible) in excess to the hydrostatic force (the material load). In (\ref{Eqn_13}), the restoring force is decreasing as a function of the wave number together with the dispersion parameter. So, $C_p$ for landslide induces an anti-restoring force. For debris material during the primarily down-slope motion this contributes positively, because this is the anti-restoring force. This is in contrast to the classical dispersive wave in string with stiffness, which is the restoring force. Due to the anti-restoring force, landslides are more dispersive than the piano strings. This reveals that the dispersion behavior in mass flow is fundamentally different than that in classical stiff-string wave motion. However, our dispersion relation, in principle, agrees with classical water waves: waves with higher wave length move faster.  

\item All the frequencies and modes of dispersion of waves in landslides can be acquired from (\ref{Eqn_7}) or (\ref{Eqn_9}) from which we may construct the sounds associated with landslides as for piano.

\item For a reasonably larger wave length the surface tension effect can be neglected. And thus, the gravity-capillary wave can well be approximated simply by the surface-gravity wave. For this reason, we have neglected the surface tension.

\item As friction and slope geometry are other important aspects in mass flows, the more complete picture of the wave dispersion in landslide can be achieved by including the additional effects of the fiction and topography (curvature) related terms in (\ref{Eqn_7}). This may result in a complex combination of restoring- and anti-restoring force regimes, possibly with the group velocity being in the direction opposite to the phase velocity. These sophisticated aspects can be dealt with separately. 

 \end{itemize} 
 
 \subsection{Implications of the dispersion relation in mass flow simulations}
 
 The above results demonstrate the importance of dispersion in legitimately simulating the wave phenomenon in naturally dispersive mass flows. The very special form of the wave frequencies, phase and group velocities and the dispersion number shown in Fig. \ref{Fig_1}, Fig. \ref{Fig_2} and Fig. \ref{Fig_3} are due to the novel dispersive wave equation (\ref{Eqn_7}), or the dispersion relation (\ref{Eqn_11}), representing the mass flow problem incorporating the effective lateral stress (normalized by mass density) $\mathcal T$, and the dispersion parameter $\mathcal D$. The overall wave dynamics are determined by $\mathcal T$ and $\mathcal D$, while dispersion is solely dependent on $\mathcal D$. The major feature of the dispersion relation is to tell us how the waves of different wave lengths move with different frequencies. So, it can play an important role in debris surge generation and attenuation. 
 
 \section{Summary}
 
 Based on the non-hydrostatic mass flow model (Pudasaini, 2022), we derived a novel dispersive wave equation, or a dispersive partial differential equation, the first of this kind, for the landslide motion. This reduces to the simple classical wave equation when the non-hydrostatic dispersion effects are ignored. Our new system of dispersive wave for debris mixture recovers the classical dispersive water waves as a special case. The frequency dispersion relation to our model for mass flow is different than the classical linear frequency dispersion for Boussinesq water wave equations. Our results show that the wave frequency with dispersion increases non-linearly as a function of the wave number. The wave frequency without and with dispersion are fundamentally different. For dispersive landslides, the wave frequency without dispersion appears to heavily overestimate the dispersive wave frequency for higher wave number. Due to the dispersion parameter emerging from the non-hydrostatic contribution for mass flow, the phase velocity becomes a function of the wave number. This gives rise to the group velocity that is significantly different from the phase velocity, characterizing the dispersive mass flows. The dispersive phase velocity and group velocity decrease non-linearly with the wave number. The dispersive group velocity is substantially lower than the phase velocity. 
 \\[3mm]
 We analytically derived the dispersion number as the ratio between the phase velocity and the group velocity. The dispersion number measures the deviation of the group velocity from the phase velocity and provides a dynamic scaling between these two velocities. The dispersion number increases rapidly as the wave number increases, and summarizes the overall effect of dispersion in the wave dynamics in mass flow. While the dispersion number for mass flow increases continuously as a quadratic function of the wave number, the dispersion number for the water waves (for intermediate depth) is strongly bounded (within the small wave number) between the shallow water and the deep water dispersion numbers. Along with other physical parameters, the mean flow depth plays an important role in determining the wave frequency, phase and group velocities, and the dispersion number. Our model and results demonstrate the importance of dispersion in legitimately describing the wave phenomenon in dispersive mass flows. 
\\[3mm]
 We defined the effective dispersive lateral stress for landslide. As the wave number increases, the effective stress decreases, however without changing the material density. Contrary to the classical dispersive wave in string with stiffness, which is associated with the restoring force, we proved the existence of an anti-restoring force in landslide. Due to the anti-restoring force, landslides are more dispersive than the piano strings. This reveals the fact that the wave dispersion in landslide is fundamentally different than the wave dispersion in the piano string. As for piano, there is now a possibility to construct the sounds associated with dispersive landslides.

\end{document}